\documentclass[%
 reprint,
 amsmath,amssymb,
 aps,
prb,
]{revtex4-2}
\usepackage[dvipsnames]{xcolor}
\usepackage{mathrsfs}
\usepackage{graphicx}
\usepackage{dcolumn}
\usepackage{bm}
\usepackage{upgreek} 

\begin{document}

\preprint{APS/123-QED}

\title{Pathways to Elastic Turbulence in Giant Micelles Through Curvature Ratios in Taylor-Couette Flow}

\author{Xiaoxiao Yang$^{1,2}$}
\author{Darius Marin$^{1}$}
\author{Charlotte Py$^{1}$}
\author{Olivier Cardoso$^{1}$}
\author{Anke Lindner$^{2}$}
\author{Sandra Lerouge$^{1}$}%
 \email{sandra.lerouge@univ-paris-diderot.fr}
\affiliation{$^{1}$ Laboratoire Matière et Systèmes Complexes, CNRS UMR 7057 Université Paris Cité, 10 rue Alice Domon et Léonie Duquet, 75013 Paris, France
}

\affiliation{$^{2}$
 Laboratoire de Physique et Mécanique des Milieux Hétèrogènes, UMR 7636, ESPCI Paris, PSL Research University, CNRS, Université Paris Cité, Sorbonne Université, Paris, France
}
\date{\today}

\begin{abstract}
In the past fifteen years, flow instabilities reminiscent of the Taylor-like instabilities driven by hoop stresses, have been observed in wormlike micelles based on surfactant molecules. In particular, purely elastic instabilities and turbulence have been shown to develop on top of shear banding, a type of flow specific to the semi-dilute and concentrated regimes. These instabilities have been identified as the origin of the large body of data showing complex spatio-temporal fluctuations, collected in shear-banded systems using multiple experimental techniques. Different categories of banding have been suggested depending on their stability, which involve intrinsic properties of the system and streamline curvature. It has been shown qualitatively that instabilities are promoted by an increase of the surfactant concentration or of the curvature of the flow geometry, while an increase in temperature stabilizes the flow. Here, using benchmark shear banding micellar systems, we quantify, for the first time, the effect of the streamline curvature on these flow instabilities, focusing more specifically on the transition towards purely elastic turbulence. Using various optical visualizations, we identify two transitional pathways to elastic turbulence. We construct a generic state diagram in a parameter space based on the curvature ratio and the Weissenberg number. The nature --supercritical \textit{vs} subcritical-- of the transition to \textcolor{black}{elastic} turbulence is discussed. \textcolor{black}{The stress evolution is in favor of  a change of nature from subcritical to supercritical transition as the curvature ratio increases. However we show that finite size effects cannot be neglected and may smooth artificially the stress response}. Furthermore, each domain of this diagram is characterized using velocimetry measurements. Finally a scaling for the onset of elastic turbulence is determined.

\end{abstract}

\maketitle

\section{\label{sec:level1}INTRODUCTION}

Giant micelles, known for their versatile rheological properties, are widely used in industrial and technological applications \cite{dreiss2007wormlike,yang2002viscoelastic,olmsted2008perspectives}. Their internal microstructure contributes to their dynamical properties and has led to their designation as "living polymers" \cite{fardin2014flows,lerouge2010shear,divoux2016shear}. These micelles exhibit viscoelastic responses, allowing them to relieve stress via dynamic breaking and recombination mechanisms, as well as undergoing reptation \cite{turner1991linear,parker2013viscoelasticity}. In the fast breaking regime, this results in a nearly perfect Maxwellian behavior for small deformations, with a single relaxation time $\tau$ and a high frequency elastic modulus $G_0$ in both semi-dilute and concentrated regimes \cite{berret2006rheology,cates2006rheology}. 

Under simple shear flow, semi-dilute micelles exhibit a weakly shear-thinning behavior at low shear rates due to progressive alignment of the micellar micro-network. Above a critical shear rate $\dot{\gamma}_1$, the flow undergoes a mechanical instability prescribed by an underlying non-monotonic constitutive relation and rearranges into two shear bands of differing internal structure, stacked along the flow gradient direction and separated by an interface of finite width~\cite{Lerouge:2020}. The signature in measured flow curves (shear stress $\sigma$ $vs.$ shear rate $\dot{\gamma}$) manifests as a stress plateau at a well-defined shear stress $\sigma_p$, where the apparent viscosity thus decreases as $\eta \sim \dot{\gamma}^{-1}$~\cite{divoux2016shear,fardin2012potential,fardin2012shear}. Recently, it has been discovered that this flow separation due to a shear banding (SB) instability is itself most often  unstable: In Taylor-Couette (TC) flow with a gap $e$, height $H$ and inner radius $R_i$, the interface between high and low shear rate bands is modulated along the vorticity direction and secondary Taylor-like vortex flows develop in the high-shear rate band~\cite{lerouge2006interface,lerouge2008interface,fardin2009taylor}. Various flow pattern were identified in the banding regime depending on the applied step shear rate, including a zig-zag pattern, anti-flame pattern, standing vortices, flame pattern or diwhirls~\cite{lerouge2008interface,fardin2012interplay}. Furthermore, under sufficiently strong velocity gradients, the presence of a disordered flow state, reminiscent of elastic turbulence (ET) \cite{groisman2000elastic,groisman2004elastic} was also observed, characterized by a non-trivial stress jump and large velocity fluctuations~\cite{fardin2010elastic,fardin2012instabilities,beaumont2013turbulent}. This phenomenology was found to be ubiquitous across giant micelles systems flowing in TC flow~\cite{fardin2016shear}. 

Semi-dilute and concentrated SB wormlike micelles are characterized by a large apparent viscosity $\eta$, even in the high shear rate band, where it is typically one thousand times  larger than the viscosity of water. Consequently, the Reynolds number, which can be expressed as $Re=\dot\gamma\tau_{\nu}$ with $\tau_{\nu}\equiv e^2/\nu$ the viscous diffusion time ($\nu$ is the kinematic viscosity and $e$ the characteristic size of the flow), is vanishingly small and inertia cannot be responsible for the secondary instability described above. Indeed, in those non-Newtonian fluids, the nonlinearity rather comes from the constitutive equation and its magnitude is given by the Weissenberg number $Wi\equiv\dot\gamma\tau$ with $\tau$ the relaxation time of the fluid. The discovery of an undulating interface \cite{lerouge2006interface} between high and low shear rate bands in TC flows has been demonstrated to be mainly due to purely elastic instability of the high-shear rate band, resulting in the presence of secondary Taylor-like vortex flows attached to this band~\cite{lerouge2008interface,fardin2009taylor,Nicolas:2012}. Note that interfacial modes due to the jump in the second-normal stress difference across the interface may also be at play in a small shear rate range, at the very beginning of the stress plateau, where the high-shear rate band remains tiny and the effect of the curvature negligible~\cite{Nicolas:2012,Fielding:2010}. This secondary instability that develops on top of the primary SB instability results from the coupling between the nonlinear elasticity of the shear-induced structures (the first normal stress difference) and the curvature of the streamlines, in a way reminiscent of purely elastic instabilities well known to develop in the homogeneous flow of polymer solutions~\cite{Larson90,Muller:2008}. In the latter case, under inner rotation and small gap limit conditions ($e\ll R_i$), the relevant dimensionless group that controls the onset of the instability between the base Couette flow and the first mode is given by $\Sigma\equiv\sqrt{\Lambda}Wi$ with $\Lambda\equiv e/R_i$ the curvature ratio, as derived analytically and observed experimentally by Larson \textit{et al.}~\cite{Larson90}. Building on the analogy with the homogeneous flow of polymer solutions, an instability criterion for SB flows has been proposed where the gap $e$ has been replaced by the size of the high-shear rate band $\alpha_he$, which acts as an effective gap, and the global Weissenberg number $Wi$ has been replaced by the local Weissenberg number $Wi_h$ in this band~\cite{fardin2011criterion}: 
\begin{equation}
	\Sigma_{SB}\equiv\sqrt{\alpha_h\Lambda}Wi_h
	\label{crit}
\end{equation}
where $\alpha_h$ is the proportion of the high-shear rate band.
Following that criterion, systems with higher local Weissenberg number or/and flowing in geometries with larger curvature ratio should be more prone to destabilization. 

\begin{figure}[!t]
	\centering
	\includegraphics[width=3.3 in]{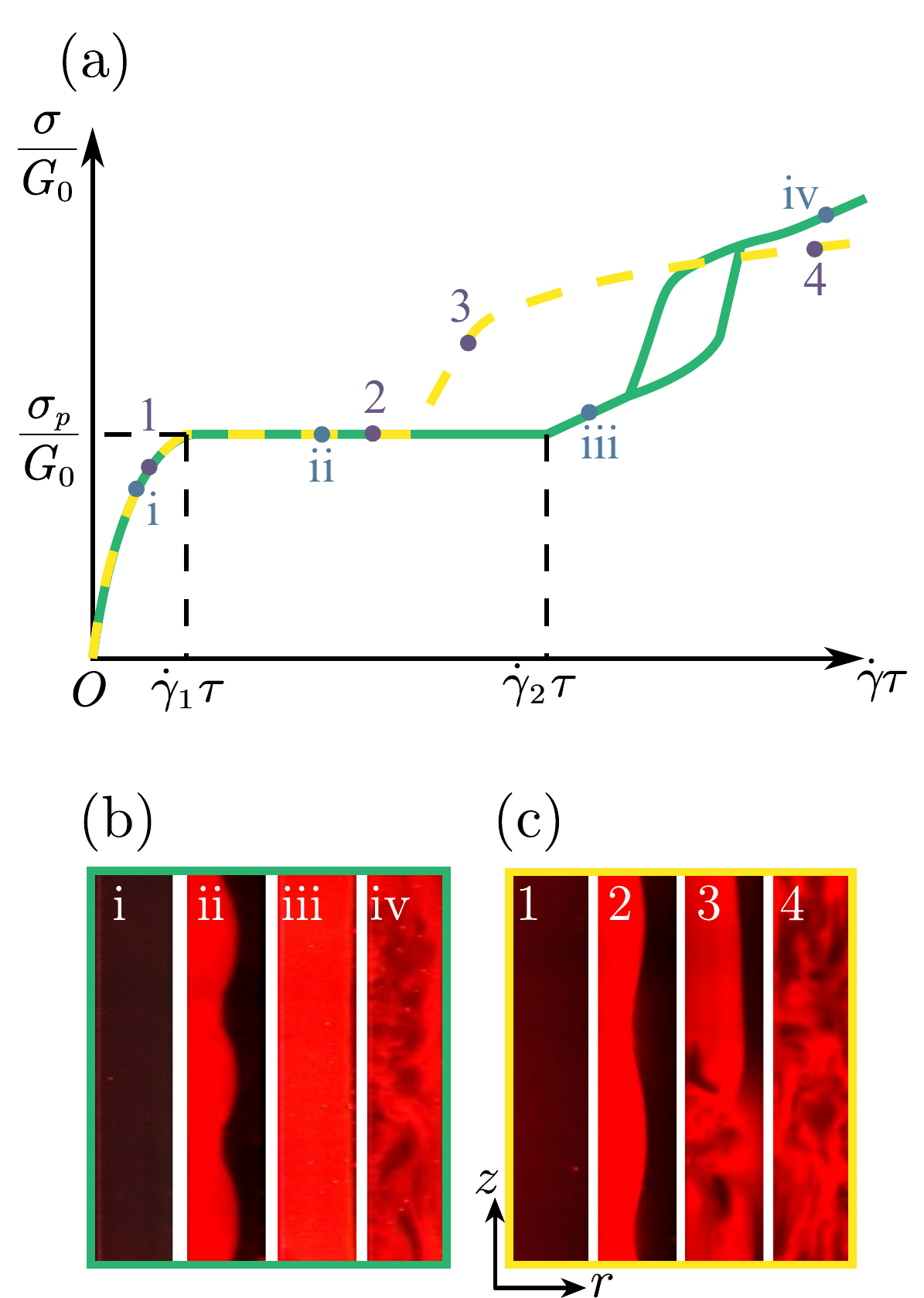}
	\caption{(a) Schematic dimensionless flow curves illustrating SB categories C2 (green) and C3 (yellow) from two shear banding micellar systems at a fixed curvature ratio $\Lambda = 0.085$ \cite{fardin2012interplay,fardin2011criterion,fardin2010elastic}. The green curve represents the flow curve of the CTAB/NaNO$_3$ system at a temperature of 28$^{\circ}$C, while the yellow curve represents the flow curve of the CPCl/NaSal system at 21.5$^{\circ}$C. Note that the effect of the cell curvature on the stress plateau has been neglected in the scheme. (b-c) Instantaneous snapshots of the banding structure in the (r, z) plane of the TC device for CTAB/NaNO$_3$ and CPCl/NaSal systems respectively, under different applied shear rates. The snapshots  correspond to distinct flow states and are marked with numbers on the flow curves in (a).}
	\label{fig:SB_category}
\end{figure}

Initially one scenario of interaction between SB and elastic instabilities was reported on the benchmark CTAB (cetyltrimethylammonium bromide) and NaNO$_3$ (sodium nitrate) system~\cite{lerouge2008interface}. A more complex interaction was revealed when investigating another benchmark micellar system made of CPCl (cetylpyridinium chloride) and NaSal (sodium salicylate) in brine~\cite{fardin2012shear}. We have reproduced these results that are illustrated in fig.~\ref{fig:SB_category}.b and c. 
Then, by qualitatively but systematically screening the effects of thermodynamical parameters (surfactant concentration, salt concentrations and temperature) and geometrical parameters (curvature ratio $\Lambda$ and aspect ratio $\Gamma\equiv H/e$), Fardin \textit{et al.} identified three categories of SB for the CTAB system~\cite{fardin2012interplay}. 
The first category (C1) corresponds to stable SB flows: in that case, the interface between the two bands remains flat and no secondary flows develop. In the general flow-phase diagram of semi-dilute and concentrated wormlike micelles~\cite{Ber97a}, the window of stability is restricted to low surfactant concentrations, close to the lower boundary of the semi-dilute regime or high temperature~\cite{fardin2012interplay}. 
The second category (C2) of SB is illustrated in fig.~\ref{fig:SB_category}.a and b. For systems belonging to this category, the SB flow is unstable all along the stress plateau. For constant applied shear rate $\dot\gamma$, shear bands first form, separated by a flat interface before secondary flows develop in the high-shear rate band after some time that depends on $\dot\gamma$. The resulting undulation of the interface profile along the vorticity direction can easily be observed in the instantaneous snapshot of the $(r,z)$ plane illuminated with laser light due to the slight intrinsic turbidity of the shear-induced structures (fig.~\ref{fig:SB_category}.b.ii). The wavelength of the pattern and the flow dynamics also strongly depend on the position of the applied shear rate along the stress plateau. Beyond $\dot\gamma_2$, the high-shear rate branch is reached, the shear-induced structures saturate the entire gap without any vortex structures and the flow homogeneity is restored (fig.~\ref{fig:SB_category}.b.iii). Thereafter, further increasing the shear rate leads to an additional bifurcation towards the fully developed \textcolor{black}{elastically} turbulent state, illustrated by large turbidity fluctuations (fig.~\ref{fig:SB_category}.b.iv) and characterized by a hysteretic stress jump (fig.~\ref{fig:SB_category}.a - green curve), typical of subcritical instabilities~\cite{morozov2005subcritical,pan2013nonlinear,morozov2007introductory,morozov2022coherent,qin2017characterizing,qin2019flow,datta2022perspectives}.

Finally, for systems belonging to the third category (C3) of SB, the phenomenology is essentially the same as in the previous case (compare for example figs.~\ref{fig:SB_category}.b.i and ii with  figs.~\ref{fig:SB_category}.c.1 and 2), except that the bifurcation towards the \textcolor{black}{elastic} turbulent state is triggered well-inside the banding regime, \textit{i.e.} before the upper boundary $\dot\gamma_2$ of the stress plateau is reached (fig.~\ref{fig:SB_category}.a and c). The banding structure together with the coherent secondary flows are disrupted by the random or regular occurrence (depending on the applied shear rate) of \textcolor{black}{elastic} turbulent bursts that also causes a significant increase of the shear stress~\cite{fardin2012shear}. These previous studies have set a clear framework regarding the stability of shear banding flows of giant micelles, in agreement with the simple dimensional argument based on Eq.~\ref{crit}. However, the effect of the curvature ratio on the transitional pathway has not been quantitatively examined. Furthermore, a scaling for the onset towards the fully developed \textcolor{black}{elastically} turbulent state is missing. 

In the present study, we quantitatively investigate for the first time the role of the curvature of the TC geometry on the stability of the shear banding flow in wormlike micelles. Using two benchmark micellar systems of the literature, CTAB/NaNO$_3$ and CPCl/NaSal, we more particularly focus on the impact of the curvature ratio on the full pathway towards elastic turbulence. Instead of the shear startup protocols employed so far, we probe the flow dynamics by performing ramping protocols in order to finely approach the critical conditions to \textcolor{black}{elastic} turbulence. As expected, we observe that the flow becomes increasingly unstable with increasing $\Lambda$. Furthermore, the shear stress evolution suggests a change of nature of the  bifurcation to \textcolor{black}{elastic} turbulence from  subcritical to supercritical when increasing $\Lambda$. However, examination of the flow dynamics shows that finite size effects cannot be neglected, indicating that a subcritical bifurcation cannot be discarded for the highest curvature ratios. We also show that, whatever the micellar system under consideration, the pathways to the fully developed \textcolor{black}{elastically} turbulent state can be recast into a generic transition state diagram in the ($\Lambda $, $Wi$) parameter space. Our findings confirm the framework established previously but they especially emphasize its generic character by showing explicitly the existence of the category C2 also in the CPCl system. Finally the geometric scaling driving the onset of fully developed elastic turbulence is determined.

This paper is organized as follows. In Section~\ref{mat}, we describe the sample composition and preparation together with the experimental setup and the various experimental techniques employed. Section~\ref{results} is dedicated to the effect of the curvature ratio on the shape of the global shear stress response and its impact at local scale on the structure of the flow and its dynamics. The different pathways to \textcolor{black}{elastic} turbulence as well as the geometric scaling are also discussed. Note that throughout the manuscript, the terms `coherent structures', `coherent secondary flows', `Taylor-like vortices' will be used indifferently to designated the same thing.

\section{\label{mat}MATERIALS AND METHODS}

\subsection{Semi-dilute micellar solutions}

In this study, we investigated two benchmark surfactant solutions, namely 0.300 M cetyltrimethylammonium bromide (CTAB) / 0.405 M sodium nitrate (NaNO$_3$) and 5.9 wt.\% (0.238~M) cetylpyridinium chloride (CPCl) / 0.51 wt.\% (0.119~M) sodium salicylate (NaSal) dissolved in distilled water, which belong to the semi-dilute concentration regime. These surfactant solutions were selected due to their widespread usage in previous studies because of the exhibition of shear-banding \cite{lerouge2010shear} in a large range of conditions, including variations of surfactant and salt concentrations and temperature \cite{fardin2012interplay}. The CTAB/NaNO$_3$ system consists of a cationic surfactant and a simple salt, and was studied at a temperature of 28$\pm 0.5^{\circ}$~C, while the CPCl/NaSal system is composed of a surfactant and a co-surfactant and was studied at 21.5$\pm 0.5^{\circ}$~C. The solutions were prepared at least one week before conducting experiments. In the preparation process, all the mixtures were stored in an oven at 35$^{\circ}$~C. They were regularly stirred for 3 days and left at rest for another 3 days.

\begin{figure}[!t]
	\centering
	\includegraphics[width=2.5 in]{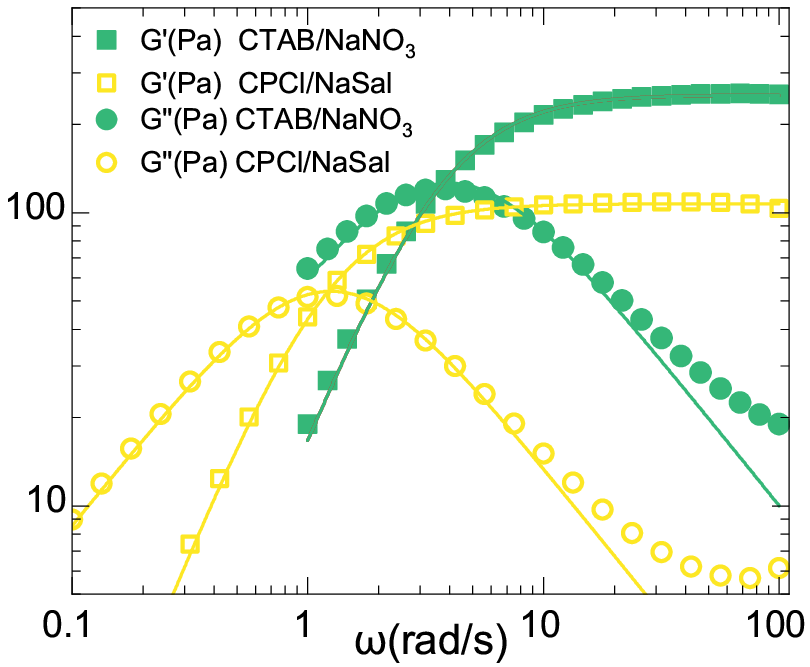}
	\caption{ Small-amplitude oscillatory shear (SAOS) linear rheology features of semi-dilute CTAB and CPCl  micellar systems. Linear viscoelastic storage modulus $G^{\prime}$ and loss modulus $G^{\prime\prime}$ are measured as a function of oscillating angular frequency $\omega$ for an imposed strain $\gamma$ = 5$\%$. The continuous solid lines indicate the best fit to the corresponding single-relaxation Maxwell model with $G_0 = 250.0 \pm 3.9$~Pa  and $\tau = 0.26 \pm 0.01$~s for CTAB/NaNO$_3$ system and with $G_0 = 102.5 \pm 6.4$~Pa  and $\tau = 0.80 \pm 0.04$~s for CPCl/NaSal system, respectively.}
	\label{fig:SAOS}
\end{figure}

 Both systems exhibit Maxwellian behavior in the linear regime, with a single relaxation time $\tau = 0.26 \pm 0.01$~s and a plateau modulus $G_0 = 250.0 \pm 3.9$~Pa for the CTAB/NaNO$_3$ system, and $\tau = 0.80 \pm 0.04$~s and $G_0 = 102.5 \pm 6.4$~Pa for the CPCl/NaSal system, as shown in Fig.~\ref{fig:SAOS}. Under simple shear flow, both systems have been known for years  to exhibit a shear-banding instability, the shear-banded state being itself unstable due to the subsequent development of elastic instability and turbulence  as previously reported in literature \cite{divoux2016shear,fardin2016shear,fardin2010elastic}.
\begin{figure*}[!t]
	\centering
	\includegraphics[scale=0.27]{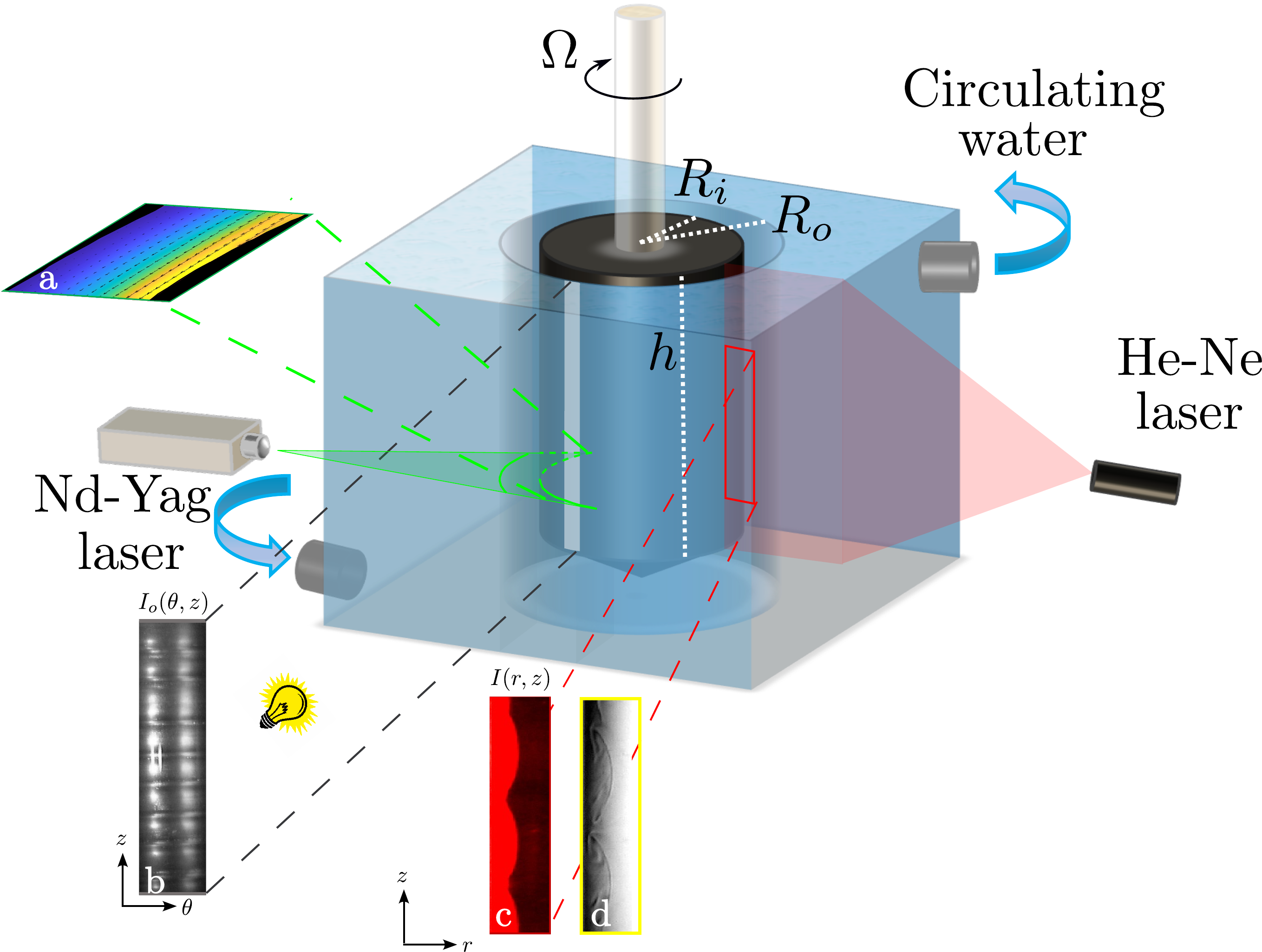}
	\caption{Sketch of the custom-designed Taylor-Couette set-up mounted vertically on a rheometer. Rheological signals can be recorded simultaneously with velocimetry images or various flow visualizations, as illustrated in insets (a) to (d).}
	\label{fig:sketch}
\end{figure*}

\subsection{Custom-designed Taylor-Couette device\label{Couette}}

Experimental investigation of the TC flow was conducted in a custom-designed set-up as shown in Fig.~\ref{fig:sketch}  mounted  vertically on a stress-controlled rheometer (Anton Paar MCR 301) operating in strain-controlled mode. The stationary outer cylinder, constructed of glass with a radius $R_o=25$~mm enclosed the rotating inner cylinder made of black Delrin$^{\mbox{\scriptsize{\textregistered}}}$. The inner cylinder attached to the rheometer shaft was of Mooney-type, the angle of the conical part being adjusted for the shear rate applied under the rotating cylinder to match that of the annular gap. The curvature ratio $\Lambda$ was varied by using inner cylinders with distinct inner radii $R_i\in \left[24,23,22,21 \right]$~mm and fixed height  $H = 40$~mm. The walls of the TC cell were smooth, and the top was closed with a metal plug to minimize fluid surface distortions and prevent bubble entrapment and sample ejection at high applied shear rates. In addition, a solvent trap was used to control evaporation. The entire TC set-up was contained in a transparent cubic envelop, and the temperature was regulated by a circulating water bath.

\subsection{Experimental parameters}

In our study, the curvature ratio $\Lambda$ was defined as the ratio between gap and radius of the inner cylinder, $\Lambda \equiv e/R_i \in \left[0.042,0.087,0.136,0.190 \right]$, where $e=R_o-R_i$ is the  gap size. As the outer radius is kept fixed the corresponding aspect ratio, defined as $\Gamma \equiv H/e$ was varied in the range $\left[40,20,13.3,10 \right]$. 
The importance of the elastic nonlinearity was expressed by the Weissenberg number $Wi$. The Reynolds number, defined in the introduction, was written as $Re = \Omega R_ie\rho /\eta$, where $\Omega$ was the angular velocity of the inner cylinder, and $\rho$ and $\eta$ were the density and the viscosity of the sample. Since the apparent viscosity decreases as $\eta \sim \dot{\gamma}^{-1}$ during the whole shear-banding regime, it was reasonable to use both the zero-shear viscosity $\eta_{0}$ and the critical viscosity $\eta_{c}$ at the onset of ET to obtain the corresponding range of Reynolds numbers [$Re_0$,$Re_c$] respectively. The largest critical $Re_c$ at the onset of ET for CTAB/NaNO$_3$ system and CPCl/NaSal system for all the curvatures were determined to be $Re_c < 0.002$ and $Re_c < 0.016$, respectively. Therefore, we can compute the corresponding range of elasticity numbers $El = Wi/Re$, which conveniently represents the relative magnitude of viscoelastic and inertial effects in a flow. It was found to be of the order of $El\in [10^{2}, 10^{4 }]$ for the two systems, indicating that the flow dynamics in the bulk of the samples were attributed to purely elastic instabilities and ET with vanishing inertia.

\subsection{Data acquisition and processing}

The flow field in the $(r,\theta)$ plane was obtained using particle image velocimetry (PIV). PIV experiments were conducted using a commercial system from Dantec Dynamics. Tracer particles with a mean diameter of 10 $\upmu$m [Silver Coated Hollow Glass Spheres (S-HGS), Dantec Dynamics] were seeded in solutions at a concentration of 100~ppm for $\Lambda \in \left[0.087,0.136,0.190 \right]$ and 120 ppm for $\Lambda = 0.042$. \textcolor{black}{We checked that the stress response of the systems was not modified by the presence of the particles~\cite{briole2021shear1}}. A flow-velocity gradient plane ($\theta, r$), approximately located 1 cm above the bottom of the inner cylinder, was illuminated using a Nd:YAG pulsed laser (DualPower 65-15 Laser, 2 $\times$ 65 mJ, $\lambda = $ 532 nm) equipped with optics generating a laser sheet with a thickness of about 0.5 mm. Pairs of images of the flow-velocity gradient plane were captured with a zoom-microscope lens (Zoom 6000 Navitar with a 1.5$\times$ objective and 2$\times$ adapter tube for $\Lambda = 0.042$, 1$\times$ adapter tube for $\Lambda = 0.087 $ and $ 0.136$, 0.67$\times$ adapter tube for $\Lambda = 0.190$) mounted on a CCD camera (FlowSense EO 4M) working at an image acquisition rate of 5~Hz below ET states and 10~Hz for ET states. The delay between laser pulses was set to vary from 15000 to 300 $\upmu$s for CTAB/NaNO$_3$ system and from 20000 to 800 $\upmu$s for CPCl/NaSal system depending on the applied shear rates. The reconstruction of the flow field was performed using the PIVlab \cite{thielicke2014pivlab} MATLAB toolbox. After subtraction of the background computed from the average of all images, the cross-correlation between two successive images in three passes of respective interrogation areas of 128, 64, and 32 pixels and with a window overlap of 50$\%$ was adopted. A typical reconstruction of the flow field in SB state is illustrated in Fig.~\ref{fig:sketch} (a).

The spatio-temporal flow dynamics were obtained using flow visualizations in the $(\theta, z)$ plane. To this end, the fluids were seeded with 1.3$\%$ of anisotropic reflective Kalliroscope AQ 1000, a suspension of anisotropic reflective platelets in propylene glycol and water, known for its effectiveness in flow-visualization in fluid dynamics. The directional deflection of the anisotropic particles with the local flow field was used to observe the kinetic evolution of the reflected light intensity when illuminated with a white light source, essentially depicting the dynamics of the axial flow structure \cite{abcha2008qualitative}. In our study, the white light source was positioned at approximately 45$^{\circ}$ to the $(\theta, z)$ plane, and a CCD camera was used to capture the reflected light intensity $I_o(\theta,z)$. The inset in Fig.~\ref{fig:sketch}(b) depicts an instantaneous image of the axial structure of the flow, where the horizontal bright stripes were attributed to Taylor-like vortices originated from elastic instability.

The evolution of the SB structure was followed using direct optical visualizations in the velocity gradient-vorticity $(r, z)$ plane. The shear-induced structures contained in the high-shear rate band  appear to be slightly turbid, differing from optical properties of the native micellar structure in the low-shear rate band. This disparity in optical properties can be leveraged to observe the banding structure and the interface between bands. To generate a laser sheet propagating along the velocity gradient axis and extending along the vorticity direction, a He-Ne laser with a wavelength of 632~nm and a cylindrical lens were used to illuminate a radial plane of the TC cell. The intensity $I(r,z)$ scattered by the sample at 90$^{\circ}$ was captured by a CCD camera to obtain a view of the gap in the $(r, z)$ plane. To follow the overall flow dynamics over the largest possible scale along the $z$ direction with reasonable spatial resolution along the $r$ direction, a CCD camera with a fixed macrolens was used. The field of observation was centered in the middle of the TC cell and varied from 7 to 20~mm based on the chosen magnification, as shown in the inset in Fig.~\ref{fig:sketch} (c). In addition, the use of a white light source to illuminate the $(r, z)$ plane in transmission using a reflecting surface behind the TC cell, akin to a Schlieren experiment, enables direct observation of the interface between the bands and gives information on the structure of the secondary Taylor-like vortex flow through the CCD camera as illustrated in Fig.~\ref{fig:sketch} (d).

\section{\label{results}RESULTS AND DISCUSSION}

\subsection{Stress response\label{rheo}}
In this section, we examine how the curvature ratio impacts the overall stress response of a shear-banding system, focusing more particularly on the mechanical signature of the transition towards the ultimate state of elastic turbulence. To avoid redundancy, we will present only the results relative to the CTAB system in the main text. Results relative to the CPCl system can be found in the Supplementary Material. 

As mentioned in section~\ref{Couette}, constrained by the outer glass cylinder that is not interchangeable, the curvature ratio has been varied by changing the radius of the inner cylinder. Consequently, the aspect ratio $\Gamma$ of the cell was also changed but remained in the limit of large aspect ratio ($\Gamma\in[10,40]\gg1$). We expect the importance of edge effects to vary with $\Gamma$. We checked that $\Gamma$  did not significantly affect the onset of \textcolor{black}{elastic} turbulence (see supplementary Figure 1), showing that edge effects are limited. 
We will see in the next section that edge effects can be detected transiently for the smallest aspect ratio.

\begin{figure}[t]
	\centering
	\includegraphics[scale=0.65]{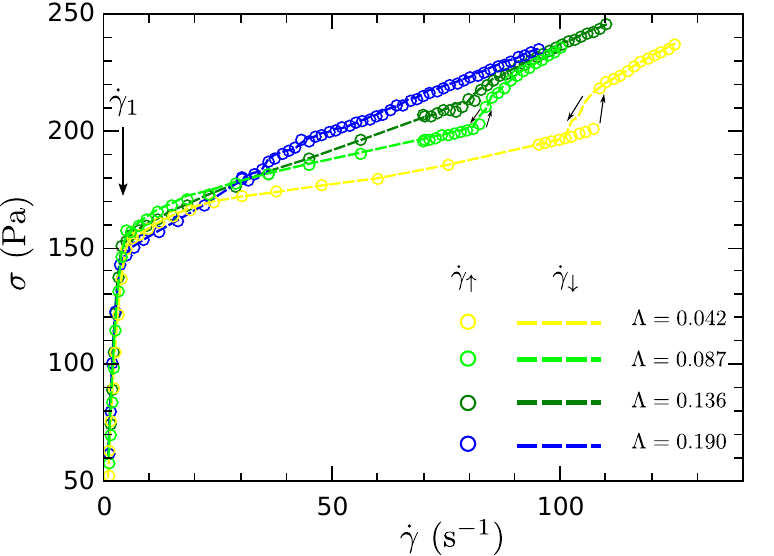}
	\caption{Steady-state shear stress $\sigma$ as a function of the applied shear rate $\dot\gamma$ measured in TC cells with different curvature ratio $\Lambda$ for the CTAB solution. The flow curves are obtained by imposing successively a stepwise increasing shear rate ramp (symbols) and a decreasing one (dashed lines). Each data point is recorded after constant shearing over 60~s. }
	\label{fig:fc}
\end{figure}

Figure~\ref{fig:fc} displays, in lin-lin scale, the flow curve $\sigma=f(\dot\gamma)$ measured in steady state conditions for various curvature ratios $\Lambda$. The shear rate was used as the control parameter in successive stepwise sweep up and down modes. For each data point, the shear stress response was recorded after a 60~s time interval, large enough for steady state to be achieved~\cite{lerouge2008interface,fardin2009taylor}. The main new finding is that the global evolution of the flow curve above the first critical shear rate $\dot\gamma=\dot\gamma_1$ is  significantly modified when changing the curvature ratio. The stress ``plateau'' is not flat and its slope notably increases with increasing $\Lambda$. It has been shown that this slope results from the combined contributions of the intrinsic curvature of the TC cell and of the 3D shear-banding flow dynamics, the secondary flows generating additional dissipation along the stress plateau~\cite{lerouge2008interface}. This point will be discussed further in section~\ref{fsd} where these two effects will be disentangled. Furthermore, as $\Lambda$ increases, the apparent extension of the stress plateau is considerably reduced: the onset of the transition towards purely elastic turbulence, which is characterized by a stress jump, is shifted to smaller values of the shear rate, meaning that the system probably  evolves from category 2 to category 3 (see fig.~\ref{fig:SB_category}) as observed qualitatively in Ref.~\cite{fardin2012interplay}. Even though the exact scaling of the onset of \textcolor{black}{elastic} turbulence is unknown, it is not unreasonable to build on what is known for the first elastic instability mode through the Pakdel-McKinley criterion~\cite{pakdel1996elastic}: increasing $\Lambda$ should drive the flow more and more unstable, the threshold being expected to scale as $\dot\gamma_c\sim\Lambda^{-\beta}$, with $\beta>0$ to be determined. \textcolor{black}{Note that in the Pakdel-McKinley criterion, which is recast in Eq.~\ref{crit} for SB fluids, the exponent $\beta$ is such that $\beta=1/2$}. In addition, in fig.~\ref{fig:fc}, as $\Lambda$ increases, the area of the hysteresis cycle is found to shrink to zero while the amplitude of the stress jump decreases, suggesting that the nature of the transition to \textcolor{black}{elastic} turbulence may change from subcritical to supercritical as the curvature ratio increases. 

Solely from the global mechanical response, it is not possible to understand the apparent changes induced by the curvature ratio. A more local description is needed. The next section is dedicated to the exploration of the flow structure and its feedback on the rheological response.

\subsection{Flow structure and dynamics\label{fsd}}

\subsubsection{Local response\label{lr}}
\begin{figure*}[p]
	\centering
	\includegraphics[scale=0.907]{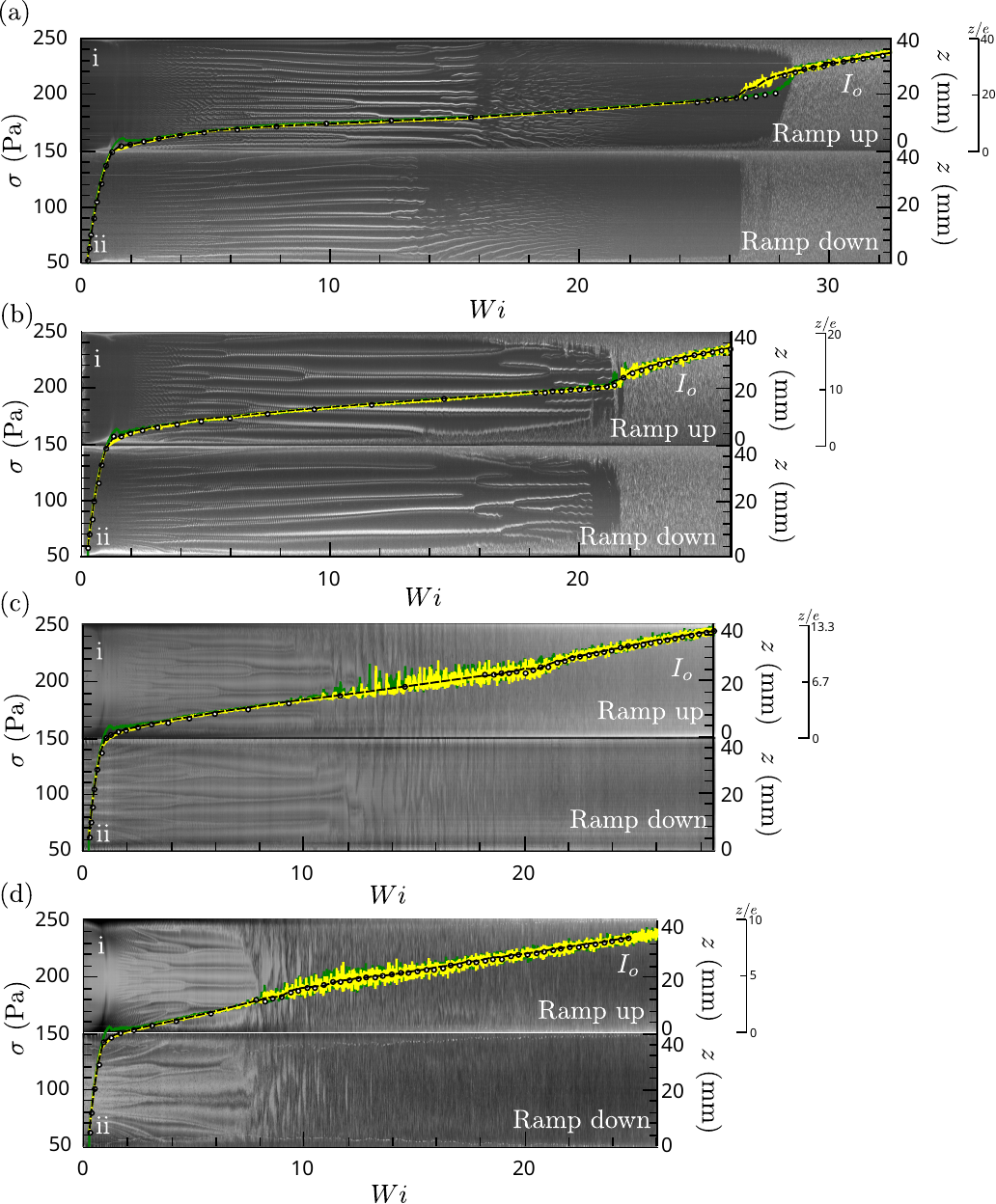}
	\caption{Shear stress response of CTAB/NaNO$_3$ system at a temperature of 28$^{\circ}$C acquired by sweeping up and down the Weissenberg number. In ``conservative'' mode, the sampling time is 0.3~s per point and ramp rate $\vert\Delta\dot\gamma/\Delta t\vert=0.054$~s$^{-2}$ (ramp up: green line, ramp down: yellow line) while in steady-stated conditions, the sampling time is 60.0 s per point (ramp up: open circles, ramp down: dash line), respectively, in the TC cell under strain-controlled mode. The combination of logarithmic and linear sampling is imposed for steady-state and linear sampling is adopted for the slow continuous ramping test. The corresponding axial spatio-temporal flow dynamics were obtained by the camera recording at 5 fps, the varying light intensity $I_o(z,t)$ reflected by anisotropic reflective particles (Kalliroscope or Iriodin) seeded in the sample. These spatio-temporal figures were made by combining a single line of pixels oriented axially along the inner cylinder of the TC cell from each frame in the complete shear rate ramp up and down video records. From (a) to (d) the curvature ratios of the geometries are $\Lambda = 0.042, 0.087, 0.136$, and 0.190, respectively.}
	\label{fig:dyn_V2}
\end{figure*}

To better resolve the flow structure and how it is modified along the flow curve, we used a convenient rheological protocol consisting in a ``quasi-static'' shear rate ramp (up and down successively) with a characteristic ramp rate chosen to ensure that acceleration-related effects are negligible and such that the measured stress follows the same path as the steady-state stress response. As 
the determination of the critical condition of an instability may depend strongly on the path used to access the state, such a type of ``conservative'' protocol turns out to be canonical for the determination of the critical conditions of Newtonian and non-Newtonian TC instabilities and their higher order transitions~\cite{Dutcher:2009,Dutcher:2013}. In our case, the ramp rate ($\vert\Delta\dot\gamma/\Delta t\vert=0.054$~s$^{-2}$) satisfied the criterion established by Dutcher and Muller~\cite{Dutcher:2009,Dutcher:2013} in purely inertial or inertio-elastic conditions $dRe_i/dt\times\tau<0.68$ where  $Re_i$ is the Reynolds number associated with the rotation of the inner cylinder and $\tau$ is the longest time scale in the system (here the micellar relaxation time). However, no equivalent exists in the purely elastic limit, which is precisely the case of shear-banding wormlike micelles. 
The ramp rate was chosen sufficiently slow for the stress response to follow the steady-state path but with the compromise that, the full test did not take too long to avoid any evolution of the sample due to evaporation problems. The shear rate increment is 0.016~s$^{-1}$ and the continuous character of the ramp allows the capture of large stress fluctuations. We tested different sampling invervals (from 0.3~s per point to 1.5~s per point) without noticing any notable change (see Supp. figure 2). 

The mechanical measurements were combined with flow visualizations along the axial direction of the TC cell through $I_o(\theta,z,t)$ and direct visualisations of the banding structure in the $(r,z)$ plane through $I(r,z,t)$. As the ramp was performed in continuous mode, the dimensionless time $t/\tau$ is equivalent to the Weissenberg number $Wi$ and the dynamics in such a mode can be represented in the $(r,Wi)$ or $(z,Wi)$ planes.

\begin{figure}[t!]
	\centering
	\includegraphics[scale=1.15]{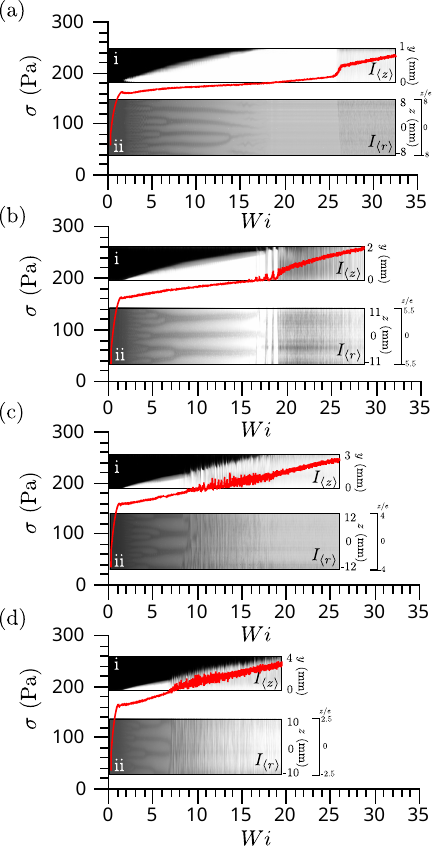}
	\caption{Correlation between the shear stress response $\sigma(Wi)$ during a quasi-static ramping protocol with increasing $Wi$ and the evolution of the banding structure tracked using the intensity $I(r,z,Wi)$ scattered by the sample. Each subplot corresponds to different curvature ratio $\Lambda=0.042$ (a), 0.087 (b), 0.136 (c) and 0.190 (d). For each one, the top diagram represents the scattered intensity $I_{\langle z\rangle}(r,Wi)$ averaged axially and provides information regarding the radial structure of the flow while the bottom diagram gives the scattered intensity $I_{\langle r\rangle}(z,Wi)$ averaged radially, which is a measure of the position $r_i(z,Wi)$ of the interface between the two bands bands (in arbitrary units) along $z$, and thus represents the rolls structure equivalently to figure~\ref{fig:dyn_V2}. The $y$ coordinate is defined such as $r=R_i+y$.}
	\label{fig:rzplane}
\end{figure}

Figure~\ref{fig:dyn_V2} displays the shear stress response $\sigma(Wi)$ to the ``conservative'' protocol with increasing curvature ratio from (a) to (d). The green and yellow lines correspond to the up and down ramps respectively while the steady-state shear stress from fig.~\ref{fig:fc} has been added for direct comparison. In each subplot, the diagrams $I_o(Wi,z)$ are represented for increasing $Wi$ [top (i)] and decreasing $Wi$ [bottom(ii)], providing information about the axial structure of the flow over the total height of the inner cylinder (the intensity has been beforehand averaged over a few pixels along $\theta$). Complementary information regarding the radial structure of the flow are provided in figure~\ref{fig:rzplane}, which is constructed in the same spirit: the mechanical response is superimposed to $I_{\langle z\rangle}(r,Wi)$ --top diagram (i)-- which can roughly be considered as the evolution of the size of the high-shear rate band (white zone) and $I_{\langle r\rangle}(z,Wi)$ --bottom diagram (ii)--, which represents the evolution (in arbitrary units) of the position of the interface between bands. Note that these two types of visualizations $I_o$ and $I$ cannot be easily performed simultaneously, the presence of anisotropic particles notably degrading the quality of the images in the $(r,z)$ plane. Consequently, from one test to another, slight differences regarding the values of both the second critical shear rate and the onset of \textcolor{black}{elastic} turbulence may arise but the overall behavior is the same and these quantitative differences do not prevent the correlation of the various dynamical events between the two views.  

Let us first focus on the smallest curvature ratio $\Lambda=0.042$ (fig.~\ref{fig:dyn_V2}.a and \ref{fig:rzplane}.a). Above a critical Weissenberg number $Wi_1 \sim 1$, the banding regime sets in, the high-shear rate band growing from the inner rotating cylinder as illustrated in fig.~\ref{fig:rzplane}.a.top diagram. Concomitantly, the formation of Taylor-like vortices are observed through the axial modulation of the reflected intensity $I_o(z)$ or the undulation of the interface profile $I_{\langle r\rangle}(z)$ (the equivalence has been demonstrated in ref.~\cite{fardin2009taylor}). As the size of the vortices was found to scale with the radial extension of the high-shear rate band~\cite{fardin2012interplay}, the wavelength of the pattern is tiny compared to the gap size at this stage. Furthermore, non-trivial periodic oscillation of the vortices along the vorticity direction ($z-$) together with localized drift towards the top or the bottom of the cell can be observed (see  Fig.~\ref{fig:dyn_V2}.a.i for $Wi=1$ to 5). Such a flow pattern is reminiscent of the rotating standing wave recently observed in numerical simulations in elastically-dominated TC flow with $El = 0.12 \sim 0.5$~\cite{lopez2022vortex}, tempting to reproduce experimental results by Lacassagne \textit{et al.} during ramp-up protocol~\cite{Lacassagne:2020}. In that specific case, the flow pattern was found to result from the interaction between vortices and a standing wave caused by the propagation of axial velocity in the outflow regions of the underlying diwhirls structure. However, in our case, despite the fact that the regime is purely elastic, the flow structure that develops, at this stage, on the top of shear banding, exhibits symmetry between the inflow and the outflow regions and does not seem to correspond to a diwhirls structure yet, which is specifically characterized by a strong asymmetry between the inflow and the outflow regions~\cite{groisman1997solitary}. 

As $Wi$ increases in the range $Wi\simeq 1$ to 14 approximately (see figs.~\ref{fig:dyn_V2}.a.i and \ref{fig:rzplane}.a.ii), the wavelength of the pattern is observed to progressively increase, similarly to the amplitude of the interface undulation, which is qualitatively given by the size of the blurred region between the white and black zones in fig.~\ref{fig:rzplane}.a.i. This evolution of the wavelength occurs throughout vortex merging events as clearly illustrated in fig.~\ref{fig:rzplane}.a.ii, as the wavelength must continuously adapt to the changing size of the effective gap. Based on the developing vortex trajectories in Figs.~\ref{fig:dyn_V2}.a.i and \ref{fig:rzplane}.a.ii, a vortex merging event shows a typical parabolic shape, which indicates that the inter-attraction of adjacent vortices strengthens and accelerates the merging process when their distance becomes comparable to their size. This behavior is consistent with the merging of two adjacent diwhirls in the TC flow of polymer solutions, as reported in~\cite{groisman1997solitary}. Consequently, the number of vortices in the gap is sequentially reduced, while their size increases significantly. Furthermore, our results indicate that the symmetry breaking between the inflow and outflow regions within a vortex pair becomes progressively more prominent with increasing $Wi$ in the range $Wi\simeq 1$ to 15. Indeed, in figs.~\ref{fig:dyn_V2}.a.i and ii, the  outflow regions (dark zones) between two inflows (bright zones) gradually widens (see also the case of $Wi = 13$ in Supp. video 1). This also suggests, by mass conservation, that the secondary radial inflow is enhanced.

From $Wi\simeq14$, the flow dynamics changes as the interface starts approaching the fixed outer wall: for $Wi$ in the range 14-16 (fig.~\ref{fig:dyn_V2}.a), the amplitude and the wavelength of the interface profile are reduced. Concomitantly, splitting of pairs of vortices separated by an outflow arises together with merging events. This behavior, also captured in fig.~\ref{fig:rzplane}.a.ii is fully compatible with the regime of creation and annihilation of pairs of vortices separated by an outflow reported in the literature for constant applied shear rates closer to the upper boundary of the stress plateau~\cite{lerouge2006interface,lerouge2008interface,fardin2012interplay,Perge:2014a}. Around $Wi\simeq16$, the flow pattern representative of the dynamics of Taylor-like vortices in the high-shear rate band seems disrupted and gives way to axially traveling waves with a much smaller wavelength (see fig.~\ref{fig:dyn_V2}.a.i for $17\leq Wi\leq22$), suggesting that the dynamics may now be driven either by the flow in the low shear rate band or by interfacial modes induced by a jump in normal stress differences at the interface between bands as at the beginning of the stress plateau~\cite{Nicolas:2012}. This ``tiny'' pattern is difficult to capture in the ($r,z$) plane (not visible in fig.~\ref{fig:rzplane}.a.ii) except using specific magnification on the outer fixed cylinder. This pattern is shown in the case of $Wi = 18.2$ in the Supp. video 1 regarding the evolution of interface between two bands via white light illumination technique. Beyond $Wi\simeq22$ in fig.~\ref{fig:dyn_V2}.a, the flow appears without any discernible pattern, suggesting that the system is now out of the banding regime, as also shown in fig.~\ref{fig:rzplane}.a.ii where the shear-induced structures fill the entire gap. In the following, this specific behavior will be referred as to ``relaminarization'' (denoted RL in the following). Over the same $Wi$ range, a very small fraction of the flow starts being strongly disordered very close to the geometric singularities represented by lower and upper edges of the rotating cylinders (see fig.~\ref{fig:dyn_V2}.a). Finally, around $Wi=Wi_c\simeq28$, the whole flow becomes turbulent leading to the expected stress jump in the rheological response. The level of stress fluctuations significantly increases along the turbulent branch. This scenario corresponds to SB of category 2 as defined in the introduction~\cite{fardin2012interplay}, where the onset of \textcolor{black}{elastic} turbulence occurs along the high-shear rate branch. As the shear rate is reduced from the fully developed \textcolor{black}{elastically} turbulent state, the turbulence persists for values of $Wi$ smaller than $Wi=Wi_c$~: the shear stress exhibits hysteresis, indicative of a subcritical bifurcation. The flow dynamics along the decreasing path is essentially the same as the one we just described. 

\begin{figure*}[!t]
	\centering
	\includegraphics[width=6 in]{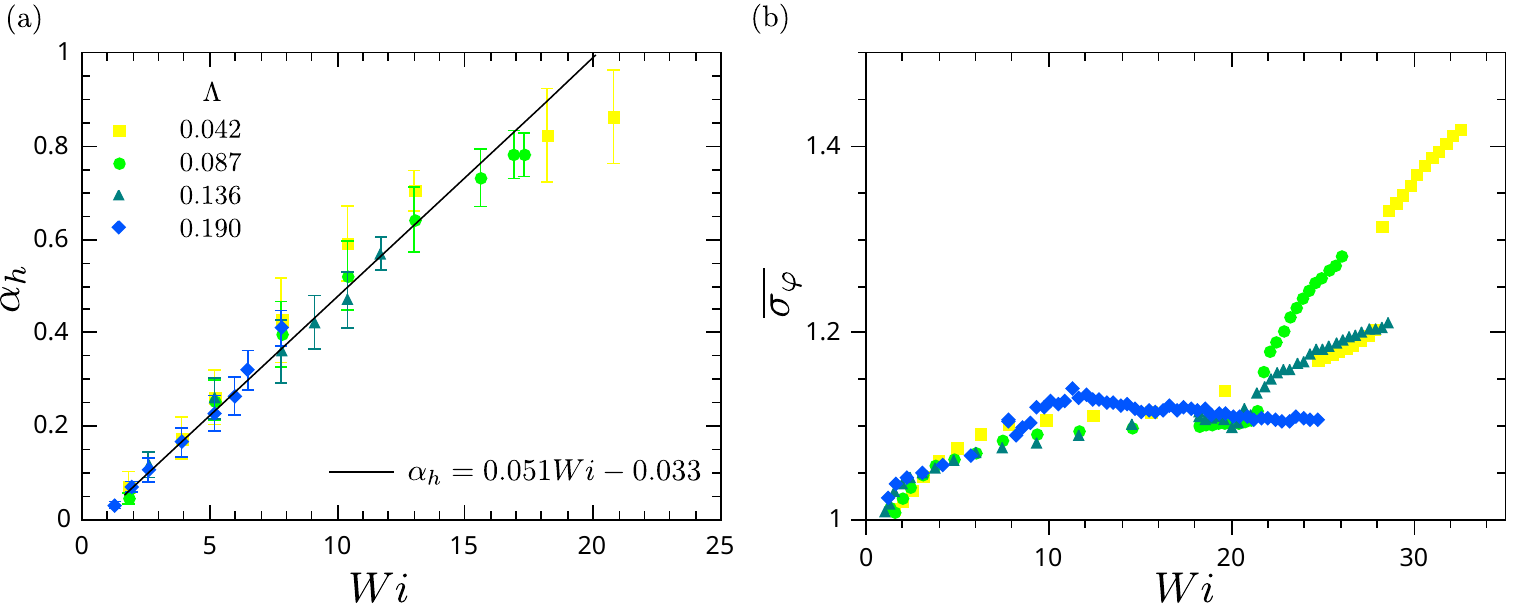}
	\caption{(a) Proportion $\alpha_h$ of the high-shear rate band as a function of $Wi$ for different curvature ratios $\Lambda$. The system is CTAB/NaNO$_3$ at a temperature of 28$^{\circ}$C. Only the SB regime is taken into account. The error bars indicates the undulating of the interface between two bands, the solid line is the linear fit based on the lever rule, $i.e.$, $\dot{\gamma}\ = \alpha_h \dot{\gamma}_h+\left( 1-\alpha_{h} \right) \dot{\gamma}_l$. (b) Dimensionless stress $\overline{\sigma_{\varphi}} \left( Wi \right)$ as a function of $Wi$ for different curvature ratios $\Lambda$.  This quantity represents the contribution of the flow resistance solely caused by elastic instabilities or ET with respect to the plateau value.}
	\label{fig:phi_ratio}
\end{figure*}

As the curvature ratio is increased to 0.087 (figs.~\ref{fig:dyn_V2}.b and ~\ref{fig:rzplane}.b.i and ii), the main trends are the following: the overall dynamics along the stress plateau is the same with successive merging of pairs of vortices as $Wi$ increases. However, at a given Weissenberg number, the number of pairs of vortices along the vorticity direction is smaller than in the previous case leading to a larger wavelength of the pattern, and earlier appearance of diwhirls with significant asymmetry between the inflow and the outflow regions. Turbulent spots that remain confined at the upper and lower edges of the inner cylinder also nucleate earlier in terms of Weissenberg number. The main difference with the previous case is that the regime of creation and annihilation of pairs of vortices is perturbed by a few turbulent bursts of finite life time propagating axially from the edges of the cell, either over the total height of the cell or over a fraction of the height. These few turbulent bursts are produced very close to the end of the banding regime as also illustrated in fig.~\ref{fig:rzplane}.b, where they span the entire gap width. They produce a large perturbation in the system and have a clear signature in the mechanical response where stress peaks are easily identified~\cite{fardin2012shear}. After the burst, the coherent structures do not reform, giving way to a tiny range where the RL state is present and the flow eventually bifurcates to the fully \textcolor{black}{elastic} turbulent state around $Wi\simeq21.5$. As mentioned in section~\ref{rheo}, the magnitude of the stress jump is reduced with respect to the previous case and the fully \textcolor{black}{elastic} turbulent state develops at smaller $Wi$ number. It is worth mentioning that at this specific curvature ratio, over the multiple tests we have performed, the number of bursts is very limited or even non-existent, leading to a more or less large range where the flow is structureless (see supplementary figure~3). In fact, the RL state is perfectly visible when decreasing the Weissenberg number (fig.~\ref{fig:dyn_V2}.b.bottom), without any turbulent burst perturbation. The comparison between ramp-up and -down protocol confirms that there is no significant difference regarding the onset of the fully \textcolor{black}{elastic} turbulent state, leading to a drastic reduction of the hysteresis at the transition. For this specific curvature ratio, the system is clearly close to the boundary between C2 and C3. 

As the curvature ratio is further increased, all trends exposed above are amplified (see figs.~\ref{fig:dyn_V2}.c and d). The number of pairs of vortices along the plateau drastically decreases for a given $Wi$. These flow structures are rapidly disrupted by turbulent bursts, which become progressively more and more frequent, giving less and less time to be fully damped until the fully \textcolor{black}{elastic} turbulent flow develops in agreement with previous observations~\cite{fardin2012shear}. For $\Lambda=0.136$ and $0.190$, the dynamics is fully dominated by the burst regime above $Wi\simeq12$ and 8 respectively, leading to fluctuations in the stress signal, much larger than those observed in the fully \textcolor{black}{elastically} turbulent state, providing a way to identify the threshold to \textcolor{black}{elastic} turbulence while the expected stress jump is barely perceptible. The views in the ($r,z$) plane show that the turbulent burst are spatially localized in the high-shear rate band and extend radially across the low shear rate band in a marginal way, essentially close to the interface between the bands (figs.~\ref{fig:rzplane}.c.i and d.i). 

All the above situations are almost exactly the same in CPCl system (see supplementary figure 4). Hence, the two systems both transition from C2 to C3 for increasing curvature ratios. 

\subsubsection{Effect on the global response\label{egr}}
Regardless of the value of the curvature ratio, the evolution of $I_{\langle z\rangle}(Wi)$ suggests that the proportion $\alpha_h$ of the high-shear rate band increases roughly linearly with $Wi$. We performed a series of startup experiments at various constant shear rates along the flow curve to get the proportion $\alpha_h$ at steady state. The results obtained for different $\Lambda$ are shown in fig.~\ref{fig:phi_ratio}(a), where $\alpha_h$ has been computed only in the range where coherent structures are present. The evolution of $\alpha_h$ follows a simple lever rule independently of $\Lambda$. In other words, whatever the value of $\Lambda$, the proportion $\alpha_h$ is the same for a given $Wi$. Consequently, the effective gap  is larger with increasing $\Lambda$ at fixed $Wi$. This is fully consistent with the observed reduction of the number of pairs of vortices with $\Lambda$ at a given $Wi$, leading to diwhirls with increasing asymmetry between the inflow and out flow regions.

To quantify the contribution of elastic instability and ET to the flow resistance, the intrinsic non-homogeneous stress variation in the gap resulting from the curvature of the geometry should be carefully considered. Indeed, the slope of the stress plateau is not only attributed to the stress distribution but also to the additional dissipation caused by the secondary flows generated by elastic instabilities~\cite{lerouge2008interface,Perge:2014a}. In the small gap limit, the stress distribution across the gap can be determined as $\sigma \left( r \right) = M /\left( 2\pi Hr^2 \right) =2R_{i}^{2}R_{o}^{2}\sigma /\left[ \left( R_{i}^{2}+R_{o}^{2} \right) r^2 \right]$ where $M$ is the torque measured by the rheometer and $\sigma$ is the shear stress measured by the rheometer~\cite{lerouge2008interface}. Therefore, the stress increment between the inner and outer cylinder walls can be given as $\varDelta \sigma \approx 2e\sigma _{p}^{*}/R_i$, where $\sigma _{p}^{*}=2R_{o}^{2}\sigma _p/\left( R_{i}^{2}+R_{o}^{2} \right) $ represents the local stress at the inner wall associated with the onset of the stress plateau. 
To disentangle the different contributions and extract only the influence of elastic instability in the slope of the plateau, we introduce a dimensionless shear stress, denoted $\overline{\sigma_{\varphi}} \left( Wi \right) =[\sigma \left( Wi \right) -\alpha _h\left( Wi \right) \Delta \sigma] /\sigma _p$. It represents the ratio between the flow resistance solely caused by the elastic instability or ET and the plateau value and is displayed in fig.~\ref{fig:phi_ratio}(b) for $Wi>1$. 

In the range of Weissenberg numbers for which the spatio-temporal response of the system is dominated by coherent structures ($Wi\lesssim8$, 12, 17, 20 for $\Lambda=0.190, 0.136, 0.087$ and $0.042$ respectively), all $\overline{\sigma_{\varphi}}$ curves collapse, suggesting that the dissipation due to the secondary flows is similar whatever the value of the curvature ratio. [$\overline{\sigma_{\varphi}}\left( Wi \right)-1$] is about 10\%, showing that the slope of the stress plateau mostly originates from the curvature ratio. However, for larger $Wi$, in the fully developed \textcolor{black}{elastically} turbulent state, the dissipation seems reduced as the curvature ratio increases.

\begin{figure*}
	\centering
	\includegraphics[width=6.6 in]{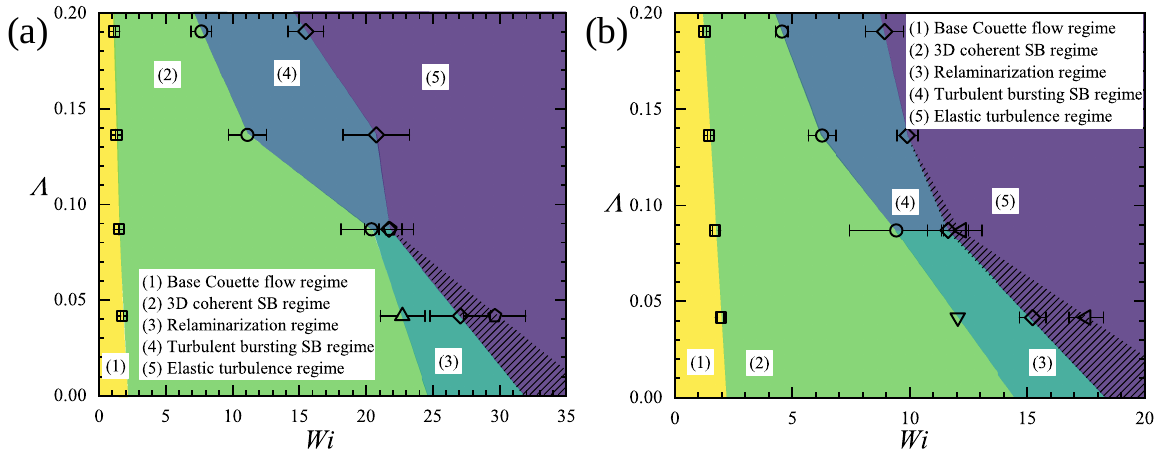}
	\caption{General state diagrams in parameter space $\Lambda-Wi$ constructed for two systems CTAB/NaNO$_3$ (a) and CPCl/NaSal (b), depending on the different $Wi$ thresholds for the distinct flow states. The evolution kinetics to the final ET state can be composed of five main regimes: (1) Base Couette flow regime, (2) 3D coherent shear banding regime, (3) Relaminarization regime, (4)  Turbulent bursting shear banding regime, (5) Elastic turbulence regime. The shading in the figure indicates the presence of hysteretic behavior. All the distinct regime thresholds are acquired from averaging over three repeated experiments. The error bar indicates the standard deviation. Therefore, two disparate transitional pathways to ET regime are identified. One is proceeding via the relaminarization to ET in which all flow structures are annihilated, and leading to a sudden shear stress jump and hysteretic loop at lower $\Lambda$. Another is featured by the occurrence of elastic turbulent bursts then eliminated by ET for larger curvature ratios.}
	\label{fig:generalsta}
\end{figure*}

Let us summarize the main findings in connection with the questions that emerged from the evolution of the global stress response as a function of the curvature ratio $\Lambda$ (section~\ref{rheo}). Despite the fact that the stress inhomogeneity intrinsic to the TC device cannot be neglected, the increase of the slope of the stress `plateau'' in the flow curve with $\Lambda$ is also due to the secondary flows, which develop on top of SB and contribute to the stress increment for about 10\% of $\sigma_p$. The overall structure of the coherent secondary flows is essentially conserved for the different curvature ratios, the aspect ratio of the Taylor-like vortices in a cross section evolving progressively towards a higher degree of asymmetry with increasing $Wi$. This asymmetry, specific to diwhirls, results from the coupling between the radial component of the secondary flow and the local hoop stress: a fluid particle starting his radial motion from the inner or the outer wall is stretched by the radial secondary flow ($\partial v_r/\partial r>0$), leading to local increase of the hoop stress, which in turn, will accelerate a particle moving inwards and slow down a particle moving outwards~\cite{groisman1997solitary}. A remarkable feature, here, is that, for a given Weissenberg number, the degree of asymmetry of the vortices is larger with increasing curvature ratio: it suggests, by a simple argument based on mass conservation, that the inward radial motion strengthens, driving the flow more and more unstable, consistently with our observations. 
Furthermore, the important changes in the stress response with $\Lambda$ (progressive disappearance of the stress jump and the associated hysteresis) are related to a change of pathway to the fully developed \textcolor{black}{elastic} turbulent state. The transition between a laminar state without any coherent structures to the \textcolor{black}{elastic} turbulent state is replaced by a regime of turbulent bursts when the system evolves  from category 2 to category 3. 

Finally, let us discuss the nature of the transition. The shrinkage of the hysteresis and the reduction of the stress jump in the flow curve with increasing $\Lambda$ suggest a change from a discontinuous (subcritical) bifurcation to a continuous (supercritical) bifurcation. However, the presence of turbulent bursts in the flow dynamics may indicate bistability between the laminar and the \textcolor{black}{elastic} turbulent states, compatible with a subcritical bifurcation. Furthermore, the bursts being produced at the edges of the inner cylinder, finite size effects may become increasingly important as $\Lambda$ increases since, at the same time, $\Gamma$ decreases. We suspect that finite size effects are responsible for smoothing the characteristics of the subcritical bifurcation to \textcolor{black}{elastic} turbulence although a change in nature towards a supercritical bifurcation is also possible. 

\subsection{General state diagram and threshold scaling}
The previous experiments have been repeated with the CPCl system and the results are shown to be completely consistent with the CTAB/NaNO$_3$ system (see supplementary figure 4). By varying $\Lambda$, our findings comprehensively suggest that there are two distinct transitional pathways to elastic turbulence in semi-dilute SB micelles, independent of the specific types of system and motivate us to propose a general state diagram to characterize these pathways, as depicted in fig.~\ref{fig:generalsta}. 
The diagram is established in the parameter space of $\Lambda$-$Wi$, revealing two disparate transitional pathways to the ET regime. The first pathway proceeds via relaminarization, where the flow structures annihilate, resulting in a subcritical transition to ET with a shear stress jump and a hysteretic loop at lower $\Lambda$. The second pathway, at higher curvature ratios, is characterized by turbulent bursts of finite life time, continuous monotonic increase of $\sigma$ without discontinuous change of slope in the flow curve. The evolution of the system can be categorized into five main regimes, as summarized in Fig. \ref{fig:generalsta}(a) and (b). The state diagrams are very similar for the two systems and recast in a generic way the effect of the curvature ratio on the stability of the SB flow. The generic character is more particularly attested by the existence of the category C2 observed for the first time for the CPCl system. 

In the following, we specify these different regimes in more details by providing the velocity profiles that characterize each of them.  

\begin{figure*}
	\centering
	\includegraphics[width=6.6 in]{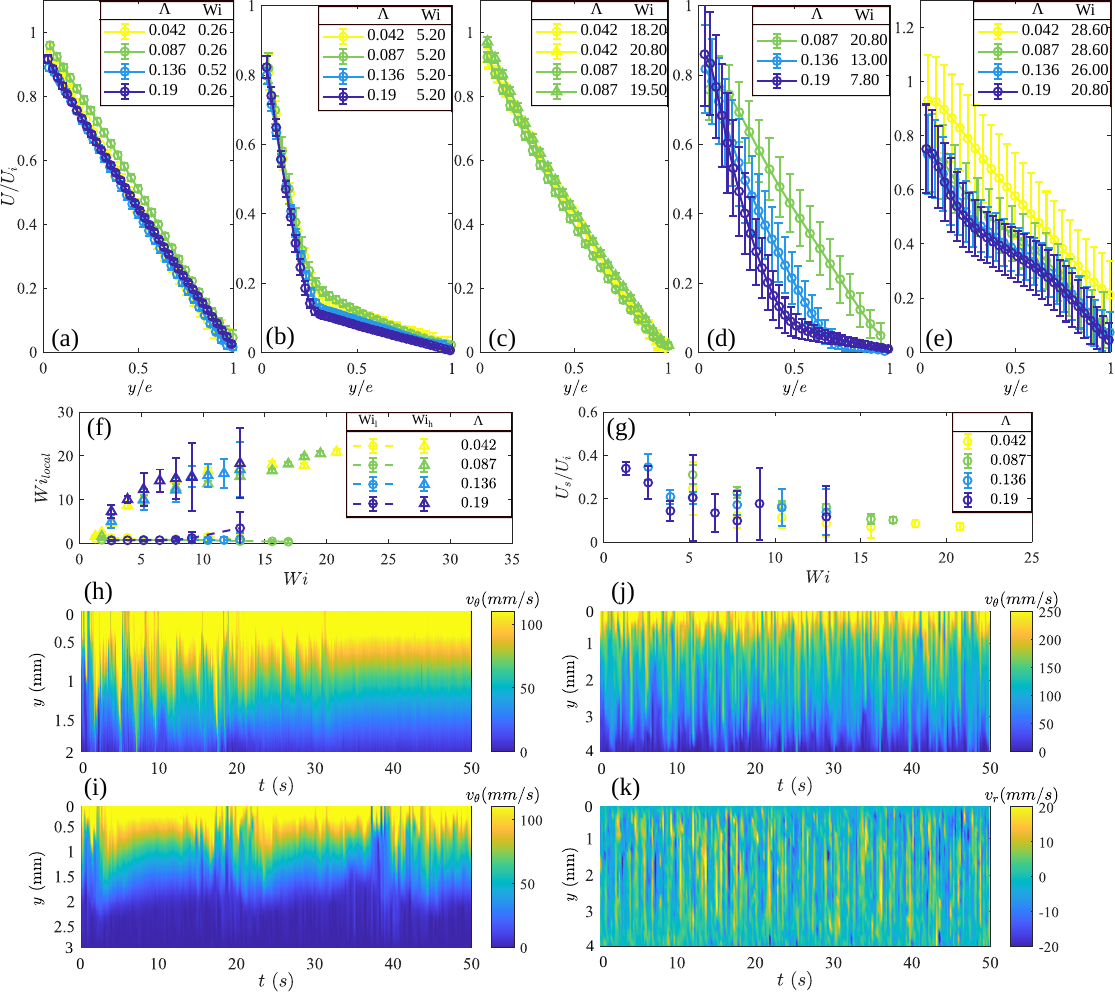}
	\caption{(a-e) 1D time-averaged velocity profiles $U/Ui$ specific of the different flow states (1) to (5) defined in figure~\ref{fig:generalsta} measured in TC flow with various curvature ratios $\Lambda$. $U_i$ is the imposed velocity at the inner wall. (a) Base Couette flow regime, (b) 3D coherent shear banding regime, (c) Relaminarization regime, (d) Turbulent bursting SB regime and (e) Elastic turbulence regime. Each profile is computed from averaging over at least 250 pairs of images except in fully turbulent regime where 500 pairs of images have been recorded. The error bars correspond to the standard deviation. (f) Local Weissenberg numbers $Wi_h$ and $Wi_l$ as a function of $\Lambda$ in the high- and low-shear rate band respectively. They are calculated by $Wi_{local} = -\tau\partial U/\partial y$, the error bars corresponding to the standard deviation. (g) Dimensionless slip velocity $U_s/U_i$ at the moving wall as a function of $Wi$. (h-i) Time evolution of $v_{\theta}(y)$ in the elastic turbulent burst regime for $\Lambda=0.087$ and 0.136 respectively. (j-k) Time evolution of $v_{\theta}(y)$ and $v_{r}(y)$ in the elastic turbulent regime for $\Lambda=0.19$. The system is CTAB/NaNO$_3$ at a temperature of 28$^{\circ}$C.}
	\label{fig:U_CTAB}
\end{figure*}

The first one, referred to as the `base Couette flow regime', corresponds to flow along the low-shear rate branch of the flow curve. As illustrated in fig.~\ref{fig:U_CTAB}.a, the flow in this domain is homogeneous and characterized, as expected, by a linear velocity profile across the gap, even for the largest curvature ratio showing that the small gap limit is respected. 

The second one, called `3D coherent shear banding regime', is associated with the plateau regime where secondary Taylor-like vortex flows stacked axially in the high-shear rate band originating from elastic instability, develop on the top of shear banding. Depending on the curvature ratio, this regime can extend either over the entire stress plateau for the smallest curvature ratios or over a variable fraction of the plateau, the apparent size of which reduces as $\Lambda$ increases. The shape of the velocity profiles in this regime is typical of the shear banding state as illustrated in fig.~\ref{fig:U_CTAB}.b. The velocity profiles are heterogeneous, with a band near the rotating inner cylinder wall ($y/e = 0$) supporting a local high-shear rate $\dot{\gamma}_h$ and a low shear rate band near the stationary outer cylinder wall ($y/e = 1$) supporting  a local low shear rate $\dot{\gamma}_l$. The level of fluctuations in the 1D velocity profiles remains weak in this regime (below 7.1\%) even if secondary flows are present~\cite{fardin2012shear,fardin2012interplay}. The local Weissenberg numbers in each band satisfy $Wi_l < Wi_e < Wi_h$, where $Wi_e$ denotes the threshold of the first elastic instability. In other words, the high-shear rate band is unstable with respect to purely elastic instability while the low-shear rate band is stable. In fig.~\ref{fig:U_CTAB}.f, we summarized the local Weissenberg numbers $Wi_l$ and $Wi_h$ determined from the measured velocity profiles for all $\Lambda$. The trends are the same as those measured previously on the same system for a fixed curvature ratio ($\Lambda=0.08$)~\cite{fardin2012instabilities}: $Wi_l$ remains close to 1 and is essentially constant in this regime while $Wi_h$ increases with $\dot\gamma$ before saturating around a value consistent with the upper boundary $\dot\gamma_2$ of the stress plateau. Interestingly, the evolution of the local Weissenberg numbers is quantitatively the same whatever the curvature ratio, consistently with the uniqueness of the lever rule determined independently  from optical measurements (Fig.~\ref{fig:phi_ratio}.a). This regime is also characterized by the development of significant slip at the moving wall while the latter was marginal in the `base Couette flow regime' (fig.~\ref{fig:U_CTAB}.g), a feature already well-known in the literature~\cite{Let2009,fardin2012instabilities,fardin2012instabilities}.

The third regime, denoted `relaminarization regime', only occurs for the lowest values of $\Lambda$ we explored and takes place as the end of the stress plateau is reached. In this regime, coherent flow structures formed previously annihilate. This regime extends over a relatively short range of applied shear rate and corresponds to the region where the high-shear rate branch is stable. Note that the existence of a range of shear rates over which the high-shear rate band becomes stable has been explained by a change of boundary conditions from the point of view of the unstable domain~\cite{fardin2012interplay}, the threshold for a soft boundary being smaller than for a hard boundary~\cite{Kha99}: indeed, as
long as $\alpha_h < 1$ the interface with the low shear rate band acts as a soft boundary while when $\alpha_h > 1$, the soft interface vanishes and is replaced by a hard wall. 
For the first time, by finely adjusting the control parameter, we unambiguously captured the homogeneity of the corresponding velocity profiles in this tiny range of shear rates (Fig.~\ref{fig:U_CTAB}.c). The profiles  exhibit moderate velocity fluctuations and slightly depart from a linear evolution, explained by the shear thinning character of the high-shear rate branch~\cite{lerouge2008interface}. The amount of slip relative to the nominal velocity appears considerably reduced in comparison with the previous regime, the local Weissenberg number being close to the global one (Fig.~\ref{fig:U_CTAB}.f). 

As the curvature ratio increases, this regime strongly reduces until it eventually vanishes, being fully replaced by a regime named `turbulent bursting SB regime', which starts at smaller Weissenberg numbers with increasing curvature ratio (see Fig.~\ref{fig:generalsta}). The case $\Lambda=0.087$ turns out to be very close to the boundary between the relaminarization regime and the elastic turbulent burst regime since the two situations can be observed from different tests with identical conditions. Consequently, for this specific curvature ratio, bursts can occur at the tail end of the stress plateau, leading to nearly linear averaged velocity profiles but with large fluctuations as illustrated by the magnitude of the standard deviation in fig.~\ref{fig:U_CTAB}.d. For larger curvature ratios, the banding structure is preserved in average, the bursts being mostly localized in the high-shear rate band. Turbulent bursts exhibit a clear signature when looking at the spatio-temporal evolution of $v_{\theta}$ at a given $z$ position, as illustrated in Figs.~\ref{fig:U_CTAB}.h and i., which display a time sequence recorded after five minutes of constant shearing for $\Lambda=0.087$ and 0.136 respectively. 
When a turbulent burst occurs, the shear banding structure is disrupted: the velocity close to the inner wall first drops, as if the low-shear rate band was intruding into the inner-wall region, before being replaced by a high velocity patch immediately afterwards as shown at $t\simeq 5$~s in fig.~\ref{fig:U_CTAB}.h or $t\simeq 40$~s in fig.~\ref{fig:U_CTAB}.i, which relaxes much more slowly towards the base banded state. Consistently with the optical observations in fig.~\ref{fig:rzplane}, the bursts span almost the whole gap for $\Lambda=0.087$ since the end of the plateau is very close while they remains confined in the high-shear rate band  for $\Lambda=0.136$. 

The final regime, called `elastic turbulence regime', corresponds to the state for which, the whole flow becomes fully disordered. Its main features are a prominent increase of the flow resistance for small $\Lambda$, together with an enhanced mixing rate (mixing in disordered and chaotic flow via laser visualization is shown in the supplemental video 2), as well as scales excited over a broad range of spatial and temporal frequencies~\cite{fardin2010elastic}. An example of the time evolution of the velocity profiles $v_{\theta}(y)$ and  $v_{r}(y)$ in this regime is provided in fig.~\ref{fig:U_CTAB}.j and k for $\Lambda=0.19$. In contrast to the 3D coherent shear banding regime where the magnitude of $v_r$ due to the vortical structures was estimated to approximate $v_r/v_{\theta} \sim \mathcal{O}(1/100)$~\cite{fardin2009taylor}, in the elastic turbulent regime, we measured that the magnitude rises to $v_r/v_{\theta} \sim \mathcal{O}(1/10)$. The corresponding profiles of the mean azimuthal velocity are reported for the first time (see fig.~\ref{fig:U_CTAB}.e). Whatever the curvature ratio, they are characterized by large fluctuations throughout the gap. Wall slip is also observed at both walls. Except the averaged velocity profile for $\Lambda=0.042$, which appears linear, probably because the corresponding applied shear rate does not go deep enough into the turbulent regime, the others exhibit an inflection point located roughly at mid-gap and present two wall regions where the local shear rate is significantly larger than in the central part. Similar shapes for the mean azimuthal velocity have been reported in direct numerical simulations of inertio-elastic turbulent states of dilute polymer solutions for larger curvature ratios~\cite{song2019correspondence}. They have also been measured in shear thickening micellar systems at high shear rates~\cite{Dehmoune:2011} where these systems are now well-know to exhibit elastically-dominated turbulence~\cite{Fardin:2014}. In the specific case of polymer solutions, the velocity profile in the central domain was much flatter.  Overall, the numerical simulations performed in the TC flow of a FENE-P fluid have shown that, the fully developed \textcolor{black}{elastic} turbulent state, both in the inertio-elastic and in the purely elastic cases, is characterized by the existence of distinct flow regions. In the purely elastic regime of dilute polymer solutions, large-scale unsteady diwhirls structures spanning across the gap mostly dominate the outer-gap region while small-scale axial and azimuthal traveling waves are predominant in the inner-wall region. The interaction between those two types of structures was found to generate fluctuating extensional flows, which, by coupling with the stretch/relaxation dynamics of polymer molecules, results in the purely elastic turbulent state~\cite{song2022direct}. It appears difficult to transpose these mechanical insights to the case of wormlike micelles but the shape of the mean azimuthal profiles observed here may also suggest the existence of spatial heterogeneity in terms of flow structures in the gap. The type of structures and their dynamics remain to be fully elucidated and is left for a future work, as well as a careful analysis of the statistical properties of the \textcolor{black}{elastic}  turbulent state.

\begin{figure}
	\centering
	\includegraphics[width=3.3 in]{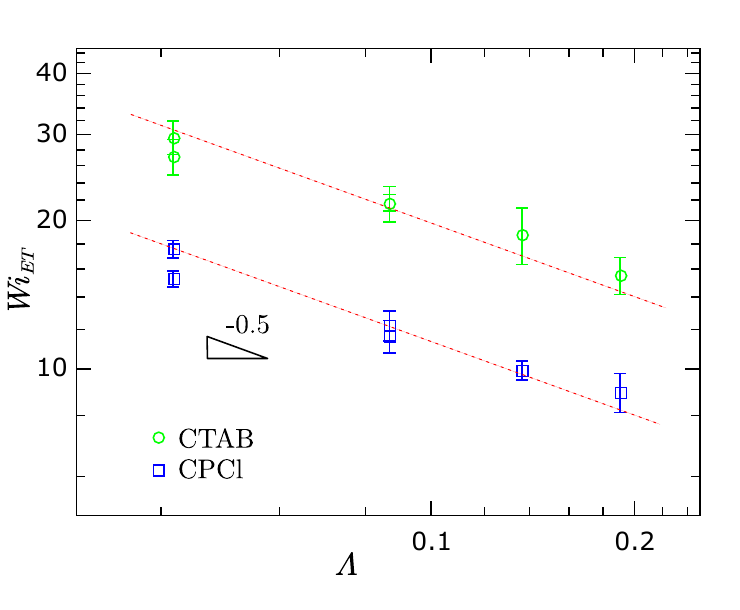}
	\caption{Onset $Wi_{ET}$  of the fully developed elastic turbulent regime as a function of curvature ratios $\Lambda$ for the two micellar systems under investigation, CTAB/NaNO$_3$ (green circles) and CPCl/NaSal (blue squares). The dotted line represents a power law function with exponent -0.5.}
	\label{fig:Wic_ET}
\end{figure}

By combining the rheological and optical measurements described in detail above, we determined the evolution of the threshold $Wi_{ET}$ to the fully developed \textcolor{black}{elastic} turbulent state as a function of the curvature ratio $\Lambda$, as illustrated in fig.~\ref{fig:Wic_ET}, for the two micellar systems investigated in this study. It is observed that the critical $Wi_{ET}$ for the transition to ET decreases as $\Lambda$ increases. Notably, both systems exhibit a power-law decay scaling $Wi_{ET} \sim \Lambda^{-{\beta}}$, with exponent $\beta\simeq 0.5$ (exponents between -0.5 and -0.4 are compatible taking into account the error bars). This geometric scaling is reminiscent of the Pakdel-McKinley scaling that recasts the different mechanisms and most unstable modes of (linear) instability of shear-dominated flow along curved streamlines, all of which involve the local enhancement of hoop stresses due to the coupling of the base state  with a velocity fluctuation, which further drives the fluctuation growth. However, here, we are considering conditions beyond the (linear) critical conditions where the subsequent elastic instabilities are subcritical and increasingly complex. Making a connection with the Pakdel-McKinley criterion remains hazardous but the scaling shows the critical role of the curvature ratio to trigger purely elastic turbulence. Note that the curves for the two systems cannot collapse along the $Wi$ axis because the dimensionless size of the stress plateau $Wi_2-Wi_1$ depends on the chosen system as shown in fig.~17 of reference~\cite{fardin2012instabilities}.

\section{CONCLUSION}

We investigated quantitatively, the effect of the curvature ratio $\Lambda$ on the flow behavior of benchmark semi-dilute shear banding micellar solutions in TC geometry in the small gap limit, focusing more specifically on the transition to elastic turbulence and the corresponding critical conditions. Increasing $\Lambda$ leads to drastic changes in the rheological response: the onset of purely elastic turbulence is shifted towards smaller applied shear rates; the significant growth of flow resistance together with its hysteretic character, which are among the main features of the transition to \textcolor{black}{elastic} turbulence, are  reduced and even disappear at sufficiently high $\Lambda$, suggesting a change in the nature of the bifurcation from subcritical to supercritical. However,  finite size effects cannot be neglected. They are intensified with increasing $\Lambda$ --as $\Gamma$ decreases at the same time-- and may also be responsible for the apparent smoothing of the transition, without any change in its nature. More work is needed to discriminate between these two hypotheses. 
Furthermore, we identified two transitional pathways to elastic turbulence depending on $\Lambda$: the first one proceeds through a relaminarization regime  at the end of the stress plateau and leads to ET above the shear banding regime and is associated with category 2 in the literature. The second one is characterized by the appearance of elastic turbulent bursts occurring while $\alpha_h < 1$ and corresponds to category 3 in the literature. On this basis, we constructed a state diagram in the $\Lambda$-$Wi$ parameter space, which appears generic across micellar systems. Each domain of this state diagram has been explored using particle image velocimetry experiments, highlighting new features. The local shear rates measured in each band are insensitive to the curvature ratio, leading to a single lever rule. Moreover, we captured the  velocity profiles in two unexplored domains of this state diagram: the relaminarization regime where the profiles appear homogeneous and slightly curved, consistently with the shear thinning behavior of the high-shear rate branch and the fully developed \textcolor{black}{elastic} turbulent state, where the profiles exhibit an inflection point with three distinct domains that may be associated with different flow structures. Finally from the combination of optical and rheological measurements, we inferred the critical conditions to the fully \textcolor{black}{elastic} turbulent state. We found that the onset $Wi_{ET}$ scales as $\Lambda^{-0.5}$, confirming the primary role of the curvature to drive elastic turbulence in TC flow. Such a scaling for the onset of the \textcolor{black}{elastic} turbulent state does not exist in the literature of regular (Boger) polymer solutions. A comparison between polymer and micellar solutions would be instructive to test its robustness. The identification of the flow structures in this state would also be particularly interesting given the large body of recent numerical simulations characterizing both purely and inertio-elastic turbulence~\cite{song2019correspondence,song2021direct,song2022direct,song2023self,song2023turbulent}. Furthermore, a full analysis of the statistical properties of the turbulent state for wormlike micelles is still lacking.

\section{ACKNOWLEDGMENTS}
The authors acknowledge funding from the PEPS CNRS program and the CSC program. They thank Marc-Antoine Fardin for the critical reading of the manuscript.

\nocite{*}

\bibliography{paperRef}

\end{document}